\documentclass[referee,a4paper,12pt,traditabstract]{jswsc} 

\usepackage{graphicx}
\usepackage{txfonts}
\usepackage{subfigure}
\usepackage{epstopdf}
\usepackage[displaymath,mathlines]{lineno}
\usepackage[authoryear,round]{natbib}
\usepackage[backref]{hyperref}
\usepackage{url}

\hypersetup{colorlinks=true,citecolor=cyan,urlcolor=cyan,linkcolor=blue}

\begin{document}


   \title{The Great Aurora of 4 February 1872 observed by Angelo Secchi in Rome}

   
   \titlerunning{The Aurora of 4 February 1872 observed in Rome}

   \authorrunning{Berrilli and Giovannelli}

   \author{F. Berrilli
          \and
          L. Giovannelli
          }

   \institute{Department of Physics, University of Rome Tor Vergata,
              Via Ricerca Scientifiica 1, I-00133 Rome\\
              \email{\href{mailto:francesco.berrilli@roma2.infn.it}{francesco.berrilli@roma2.infn.it}}\\
             \email{\href{mailto:luca.giovannelli@roma2.infn.it}{luca.giovannelli@roma2.infn.it}}
             }


 
  \abstract
   {Observation of auroras at low latitudes is an extremely rare event typically associated with major magnetic storms due to intense Earth-directed Coronal Mass Ejections. Since these energetic events represent one of the most important components of space weather their study is of paramount importance to understand the Sun-Earth connection. Due to the rarity of these events, being able to access all available information for the few cases studied is equally important. Especially if we refer to historical periods in which current accurate observations from ground-based instruments or from space were not available.\\
   Certainly, among these events we must include the great aurora of February 4, 1872. An event whose effects have been observed in different regions of the Earth. What we could consider today a global event, especially for its effects on the communication systems of the time, such as the transatlantic cable that allowed a connection between the United States and Europe since 1866.\\
   In this paper we describe the main results of the observations and studies carried out by Angelo Secchi at the Observatory of the Roman College and described in his \textit{Memoria sull’Aurora Elettrica del 4 Febbraio 1872} for the Notes of the Pontifical Academy of new Lincei. This note is extremely modern both in its multi-instrumental approach to the study of these phenomena and in its association between solar-terrestrial connection and technological infrastructures on the Earth. The Secchi's note definitely represents the first example of analysis and study of an event on a global scale, such as the Atlantic cable, affecting the Earth. What we nowadays call an extreme space weather event.
   }        


   \maketitle

\section{Introduction}
In recent decades, rapid growth in the realization of heliophysical ground-based instrument networks and space missions has contributed to an awareness of the complexity of the Sun-Earth system and of the possible impact that extreme space weather events can have on our technological society. Uncommon extreme events, such as those associated with the major auroras of the last two hundred years, 
represent case studies of capital scientific and historical interest.
Among those, the August 28 and September 1-2, 1859, and the February 4, 1872 events are extremely relevant \citep{Silverman2006}.\\
Although this awareness is recent, the systematic observation of solar activity in connection with the Earth dates back to two centuries ago. As recently discussed in \cite{Cade2015}, expressions such as \emph{solar meteorology\/}, \emph{cosmic meteorology\/} or \emph{magnetic weather\/} indicated concepts such as the appearance of structures (e.g., sunspots, prominences, etc.) on the Sun, sudden variations of the Earth's magnetic field, or events that correlate the status of the Sun with the terrestrial or circumterrestrial physical state as early as the XIX century.\\
Nowadays, with the expression \emph{Space Weather\/} \citep[e.g.][]{Lilensten2009}
we typically refer to that class of processes and physical conditions of the Sun, produced by its magnetic activity, which generate: flares, solar wind variations, especially in connection with streamers of fast and slow wind, Coronal Mass Ejections (CMEs), variable flux of charged solar energetic particles (SEPs) and which impact the physical state of the Earth's magnetosphere, ionosphere and thermosphere  \citep[e.g.][]{Alberti2018, Spogli2019, Bigazzi2020, Ward2021} or of the whole planetary system \citep[e.g.][]{Plainaki2016}. Of particular relevance to our society are those processes that can affect the performance and reliability of space and ground technology systems, including the possible health effects of astronauts or air crews \citep[e.g.][]{schwenn2006,Berrilli2014,Difino2014,Plainaki2020}.\\
This paper deals with the massive solar storm and great aurora of February 4, 1872  \citep[e.g.][]{Fawcett1872, Slatter1872, JMH1872, Stone1872, Toynbee1873, Silverman2006, Hayakawa2018, Valach2019, Berrilli2020, Oliveira2020}. In particular, we will describe the observations and the study carried out by Angelo Secchi, and supported by his collaborators, presented in the note \emph{Memoria sull’Aurora Elettrica del 4 Febbraio 1872} for the Notes of the Pontifical Academy of new Lincei on February 18, 1872 \citep{Secchi1872}. We will discuss how modern his multi-instrumental scientific approach is, today we would say that it is a \emph{multi messenger} study, and above all how the author introduced and discussed the effects on terrestrial infrastructures. In our opinion it is extremely interesting, both from a historical and scientific point of view, to present Secchi's note which for the first time connects, in a multi-instrument study, solar sources, auroral effects and the failures of technological infrastructures, like the Atlantic rope, on a planetary scale.

\section{Angelo Secchi and the Observatory of the Roman College}
As we reported in the previous paragraph, the great aurora of February 4, 1872 was one, if not the most intense geomagnetic event of solar origin of the last two hundred years. This phenomenon has been extensively described in some papers \citep[e.g.][]{Silverman2001,Silverman2008} and in a series of short reports published in the journal Nature in 1872, close to the event \citep[e.g.][]{Fawcett1872, Slatter1872, JMH1872, Stone1872, Toynbee1873} to which we refer for a broad and exhaustive description of what happened in those days of February 1872. 
\cite{Silverman2008} hypothesize that the aurora of February 4, 1872 is associated with a geomagnetic storm of greater intensity than the more famous great auroral storms during the Carrington Event in 1859 \citep[e.g.][]{Hayakawa2019}. In support they argue that the great aurora of 1872 was seen all over the world, from the Caribbean to Egypt, to the Indian Ocean and the Indian subcontinent and China, with observations extending down to $20^o$ of magnetic latitude.
An extensive work by \cite{Hayakawa2018} reexamined all the available observations of the aurora from East Asia, including Japan, Korea and China. In particular, an observation from the Italian consulate in Shanghai ($19^o.9$ magnetic latitude) reported by Donati was used to estimate the equatorward boundary of the auroral display and consequently a Dst value of about -1900 nT. The 1872 event therefore has a comparable intensity of geomagnetic disturbance and auroral extension as the 1859 Carrington event.\\
However, \cite{Silverman2001,Silverman2008} did not present the results of Secchi's note, probably because they were not acquainted with it.
In this work we focus on the observations and measurements made by the staff of the Observatory of the Roman College and on the connections and conclusions that Secchi reports within his note.

\subsection{Father Angelo Secchi}
Before describing the observatory of the Roman College and the main instruments inside, we briefly introduce the figure of Angelo Secchi, referring any further information to dedicated papers \citep[e.g.][]{Chinnici2017, Orchiston2020, Chinnici2021}. \\
Angelo Secchi was born in Reggio Emilia on June 29, 1818. In the same city he attended the Jesuit College. He moved to Rome in 1833 where, at the age of fifteen, he entered the Jesuit novitiate in Rome and then the Roman College where he distinguished himself in a course that included physics and mathematics. In 1841 he was appointed Jesuit college instructor in Loreto, but between 1844 and 1848 he returned to his theological studies. However, he did not abandon his scientific interests by collaborating with Francesco de Vico, director of the observatory and professor of astronomy at the Gregorian University of Rome.\\
In November 1848, Pope Pius IX was forced to take refuge in Gaeta and during the Roman Republic several monasteries and churches were occupied. Many Jesuits fled Rome and Father Secchi moved to the Jesuit college at Stonyhurst, Lancashire in England. Subsequently, he embarked for the United States where he settled at Georgetown University in Washington. Here, in addition to dealing with astronomy as assistant to director P. Curley, he met the hydrographer M. F. Maury and became interested in meteorology. These encounters were extremely important in expanding Secchi's scientific interests and would have had important consequences in the design of the new observatory in Rome, where he returned at the end of 1849, becoming director of the observatory of the Roman College in 1850. \\

\subsection{The observatory of the Roman College and its instruments}
In his role as director, Angelo Secchi began the design and construction of the new observatory of the Roman College. Details of the long process to design and build the observatory can be read in the beautiful memoir, written by Secchi, and entitled \textit{L'Astronomia in Roma nel Pontificato di Pio IX} \citep{Secchi1877}. 
In this memoir the author reports the important circumstance for which the great dome, 17~m in diameter and 80~m height above the floor, initially planned over the church of St. Ignazio of Loyola, was never built.
Therefore, the mighty pillars sized to support this great dome were available and were perfect to host the various telescopes and the magnetic and meteorological laboratories with the necessary stability\\
It is worth noting that a lot of instrumentation, especially for meteorological observations, was automated and with pen recorders. This automation was absolutely avant-garde for the time and in this regard it is historically interesting to report a sentence reported in the paper written on the occasion of the first centenary of the birth of Secchi \citep{Rigge1918} which presents a certain amount of intellectual prejudice towards Italians. In fact, the author wrote \textit{Father Secchi with a true Yankee ingenuity which we Americans would hardly expect to find in an Italian, set to work to make all his instruments automatic and self-recording}.
\begin{figure}
\centering
\includegraphics[width=16cm,height=11cm]{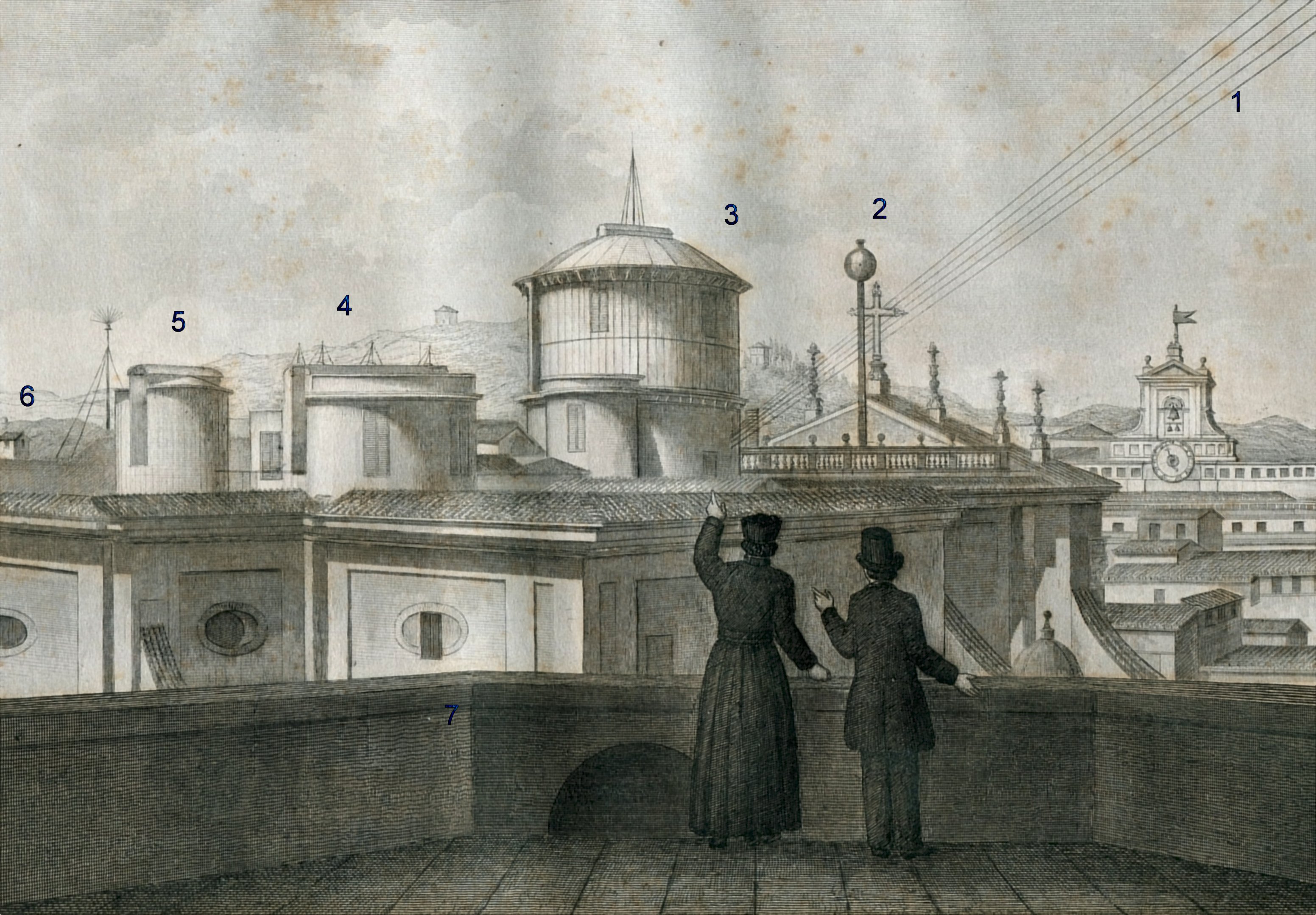}
\caption{Observatory of the Roman College seen from the Calandrelli Tower. \textbf{The reproduction shows in order}: 1) bundle of electric cables for the transmission of the signals of the meteorological sensors located on the Calandrelli Tower to the Secchi's automatic Meteorograph; 2) pole with wicker ball, used, starting from 1847, to signal midday to the artillerymen of Castel S. Angelo; 3) dome with the Merz telescope; 4) elliptical observatory for the meridian circle of Ertel; 5) dome with the Cauchoix telescope; 6) electric turret observatory with the small ball conductor; 7) Median Terrace of the Calandrelli Tower.
From Secchi's communication \textit{L’astronomia in Roma nel pontificato DI Pio IX} \citep{Secchi1877}.
}
\label{fig:observatory}       			  
\end{figure}
The Fig.~\ref{fig:observatory} shows a view of the external structures of the observatory of the Roman College seen from the Calandrelli Tower, the tower that housed the meteorological sensors.
\subsubsection{The refractor telescopes of Merz and Cauchoix}
\begin{figure}
\centering
\includegraphics[width=12.5cm,height=14cm]{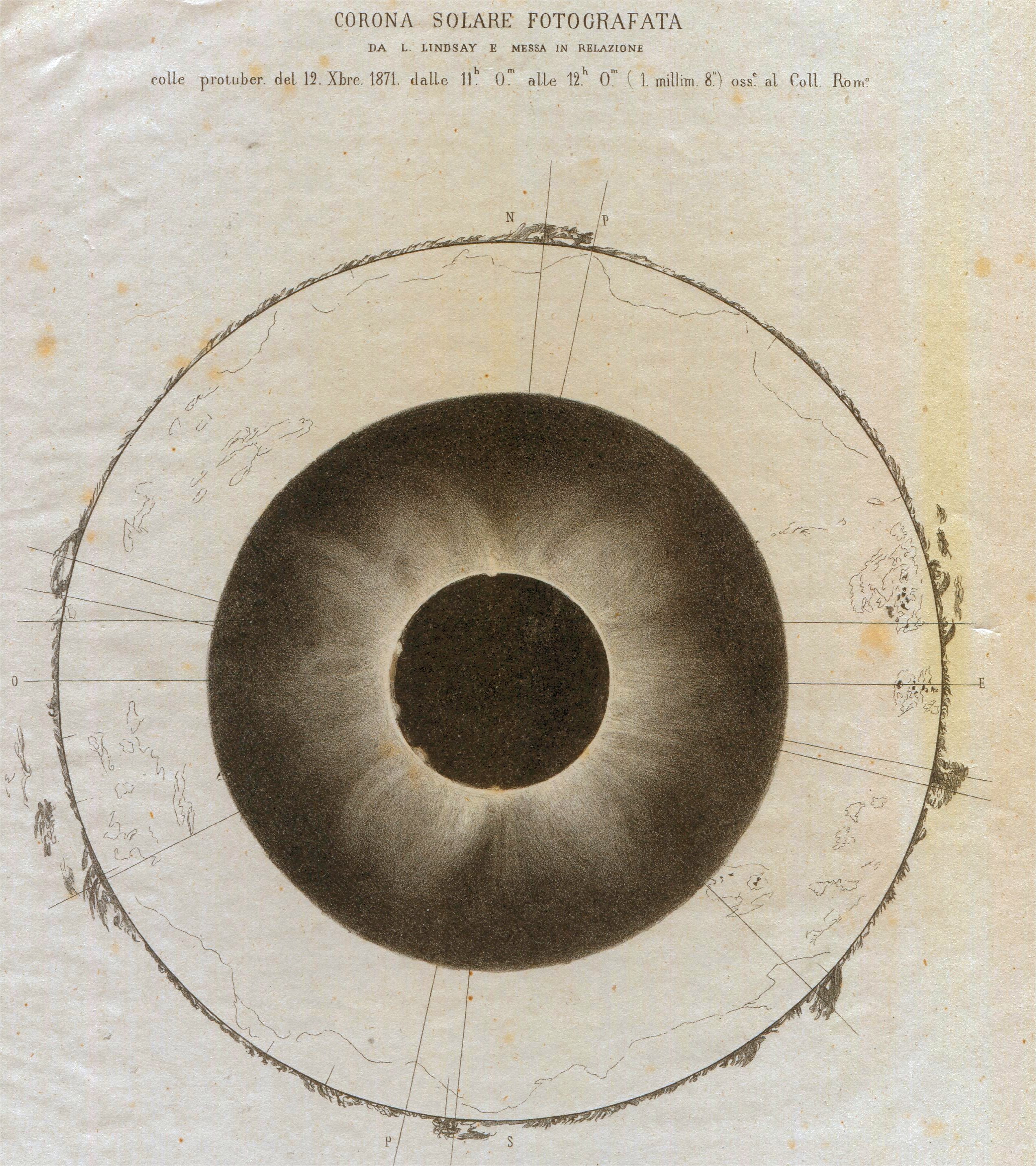}
\caption{The figure shows the photograph of the total eclipse of 12 December 1871 taken by Lindsay in comparison with the chromosphere protuberances observed from 11:00 to 12:00 by the Roman College. From Secchi's communication \textit{Sulla distribuzione delle protuberanze solari e loro relazione colle macchie} \citep{SecchiProt1873}.
}
\label{fig:corona}       			  
\end{figure}

The major telescopes for astronomical and solar observations were the Merz and Cauchoix refractor telescopes. Since Secchi wished to study the Sun spectroscopically, he commissioned Hofmann and Merz to build spectroscopes that incorporated a series of prisms. The direct vision prism was made by Giovanni Battista Amici instead.\\
The Merz telescope was a 4.3~m long free-aperture 9-inches refractor. It was installed inside the main dome (Fig.~\ref{fig:observatory},3) which Secchi called  \emph{cielo mobile maggiore\/}. This dome was 7.25m high in the centre and rotated on 8 cannonballs placed between circular channels in cast iron. The instrument was equipped with a solar reflecting prism diagonal eyepiece and two polarizing eyepieces, these were 
donated to Secchi by Warren de la Rue, Georg Merz and the inventor P. Cavalleri. In addition, the telescope had a finder scope of about 8 cm of aperture and a triple series of eyepieces: positive, for micrometric use, and negative.\\
The Cauchoix's telescope was instead installed in a smaller dome (Fig.~\ref{fig:observatory},5) called  \emph{cielo mobile minore\/}. The telescope had an aperture of 6-inches and a focal length of 238cm and was equipped with an objective prism spectroscope. It was mainly used for solar observations. These were carried out by projecting the image of the sun, which had a diameter of 24.3 cm with a focal plane scale of about 7.5~$arcsec/mm^{-1}$, onto a flat sheet of paper. 
A daily log of sunspots observed since 1857 was kept. These cover almost the entire Solar Cycle 10 which began in December 1855.\\
Describing Secchi's work in the solar field \citep[e.g.][]{SecchiSoleil1875} is beyond the scope of this paper. However his search for connections between different layers of the solar atmosphere, combining observations from different instruments, must be emphasized for his modern approach.\\
For example in the communication  \textit{Sulla distribuzione delle protuberanze solari e loro relazione colle macchie} \citep{SecchiProt1873} Secchi argues on the shape of the corona, photographed during the eclipse of December 12, 1871 by Devis and commissioned by Lord Lindsay, in connection with the observation of the chromospheric prominences observed, on the same day, by the telescopes of the observatory of the Roman College (Fig.~\ref{fig:corona}). 
Secchi comments that \emph{the coronal structure does not show straight jets or plumes, but rather curvilinear. The latter show symmetrical curvatures all in relation to the solar equator. For this I came to an important conclusion. This was that the solar atmosphere represented by this corona, it could not be considered as an equilibrium layer arranged with static rules, but which instead was constituted in a rigorously dynamic structure\/}".
\subsubsection{The magnetic and electric laboratories}
As we have seen in the previous paragraphs, Secchi developed during the years spent at Georgetown University a strong interest in magnetic and electrical observations in the geophysical field. He therefore equipped the observatory of the Roman College also with a laboratory for magnetic measurements, the first Italian magnetic observatory (more information on the subject can be found in the paper \cite{Ptitsyna2012}. The measurements made in this laboratory are those used to describe the event of February 4, 1872.\\
The routine measurements with the magnetometers were carried out 8 times a day at fixed times, between 7:00~a.m. and 9:00~p.m., simultaneously with the meteorological measurements.
All the magnetic instruments were located in a specially made room, from where all the iron carpentry was removed and replaced with wood and copper, and placed over the western arm of the Church in order to ensure its stability  (Fig.~\ref{fig:observatory},6). \\
The laboratory contained three differential magnetic instruments capable of measuring: $i)$ the variations of the earth's magnetic field declination in the South-North direction, i.e., the angle between the direction of the magnetic needle and the meridian of the site, with the declinometer; $ii)$ the variations of the Earth's magnetic field in the East-West direction by means of the bifilar magnetometer; $iii)$ the vertical magnetic component by means of the balance magnetometer. \\
Other three instruments allowed to carry out measurements of the absolute declination of the magnetic needle, the absolute inclination and the absolute intensity with the Gauss method. Secchi reports that all the instruments were built in England and that the measurement procedure basically followed the methodologies adopted by the English magnetic observers with some improvements introduced at the Roman College.\\
The observatory was also equipped with instruments for geophysical electrical measurements. These measurements were carried out with different instruments depending on the weather conditions. In the presence of thunderstorms, a fixed conductor was used. If the weather was good and the sky was clear, a Palmieri mobile conductor was used for ordinary electricity. Finally, a telegraph wire about 40~miles (about 75~km) long was used that connected Rome to Porto d'Anzio. Secchi points out that the Collegio Romano observatory was the first to carry out this type of measurements which were later adopted by other observatories including the Greenwich observatory in England.
\begin{figure}
\centering
\includegraphics[width=14cm,height=12cm]{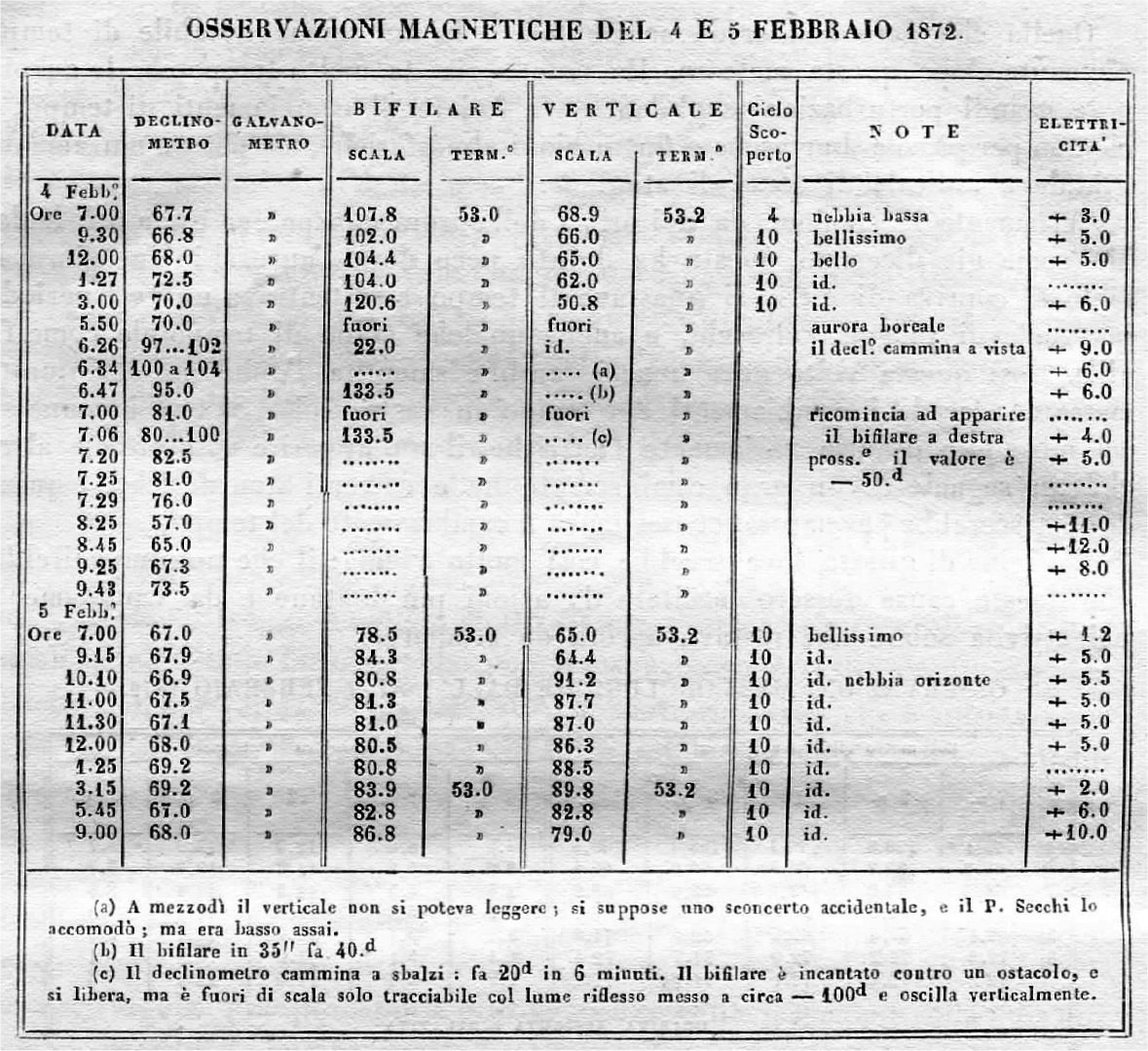}
\caption{Table of magnetic and electrical observations of 4 and 5 February 1872 carried out in the laboratories of the Roman College. The different columns record: the date and time (Rome solar time), the magnetic declination, the electric current measured by a galvanometer, the variations of the horizontal and vertical components of the magnetic field (and relative instrument temperatures), sky coverage and general notes, electrical measurements. From the note \textit{Memoria sull’Aurora Elettrica del 4 Febbraio 1872} \citep{Secchi1872}.
}
\label{magnetic_table}       			  
\end{figure}
\subsubsection{The meteorological observatory}
The interest in meteorology developed by Secchi during his stay in the United States meant that during the re-foundation of the observatory of the Roman College, modern meteorological instruments were also purchased for the study of the climate in Rome. The meteorological observatory was equipped with various instruments: e.g., rain gauge and anemometer, a Fortin barometer and psychrometer thermometers. Furthermore, the \textit{Meteorografo} was designed and built by Secchi. This complex and totally automated instrument 
recorded on sheets of lined paper the curves of the barometer, of the dry and wet thermometer (for determining the relative and absolute humidity of the air), of the speed and direction of the wind, and finally of the hour and duration of the rain. Observations were made 8 times a day, from 7am to 9pm, including a measurement, requested by US colleagues, at 1:30~pm. For more details on Secchi's \textit{Meteorografo} read the essay \citep{Brenni1993}.
\section{The observations of the aurora from the Observatory of the Roman College}
The Secchi's note on February 18, 1872 \citep{Secchi1872} opens with these words: \emph{the superb electric aurora that we witnessed on the 4th of this month, is so extraordinary, for our climates (i.e., latitudes), that it deserves to be handed down to posterity with all the particularities that were reported during his appearance; and this is more necessary since various important facts for the theory of the phenomenon have emerged\/}.\\
At 5:45~pm on February 4, 1872, Secchi's assistant F. Marchetti, while carrying out the planned magnetic measurements, noticed that the magnetometers were unusually disturbed. In particular, both the bifilar and vertical magnetometers were out of range. Observing the sky from the electrical turret (Fig.~\ref{fig:observatory},6), from which there was a complete view of the sky, he saw the aurora in direction N and N-E and immediately informed the director. 
The note contains a long and very detailed description of the phenomenon and what was observed during the night which, for reasons of space, we do not report in this work. 
The description of the aurora evolution is very accurate and describes the shapes observed, reporting their geographical or relative position to the brightest stars (e.g., Capella or Aldebaran) and the most important constellations (e.g., Orion or Gemini). It also describes the evolution of its colours, in which yellow-greenish is visible but deep red dominate at the top. The predominant red colour coincides with what is reported in \cite{Silverman2008}  where it is written that the predominant impression of colour was universally that of a rosy red or blood-red.\\
It is worth to note that while the visual and spectroscopic observations of the aurora were carried out, the magnetometers were also often consulted.
They continued to be extremely disturbed and often out of scale, only the declinometer was able to follow the entire evolution of the event (see original table reported in Fig.~\ref{magnetic_table}).\\
The following is a list of multi-instrument observations and conclusions, grouped into broad arguments, reported in Secchi's note.
\subsection{Spectral observations}
On the basis of the spectral observations, obviously imperfect due to the low luminosity of the phenomenon, Secchi noted several facts. For example, a yellowish-green line was visible everywhere, which we now know is produced by oxygen atoms located at heights below 150 km \citep[e.g.][]{Lanchester2009}.
A red line was clearly visible only in the \textit{red columns} of the aurora. This colouring, which we have said to have been noticed by many observers, is not particularly frequent and is typically associated to the forbidden oxygen red line, at 630.0 nm, indicating large fluxes of low-energy electrons
 \citep[e.g.][]{Silverman2008}.\\
Finally Secchi reports the observation of a spectrum composed of many lines, in the brightest regions of the aurora, and notes that \textit{at certain moments seemed to see a piece of the nitrogen spectrum, but it was impossible to study it because it varied instantly}. Nowadays we know that this is possible, because at an altitude of about 120 km the solar particles interact with the atmospheric nitrogen creating blue/crimson red colours.
\subsection{Magnetic observations}
\begin{figure}
\centering
\includegraphics[width=16cm,height=11cm]{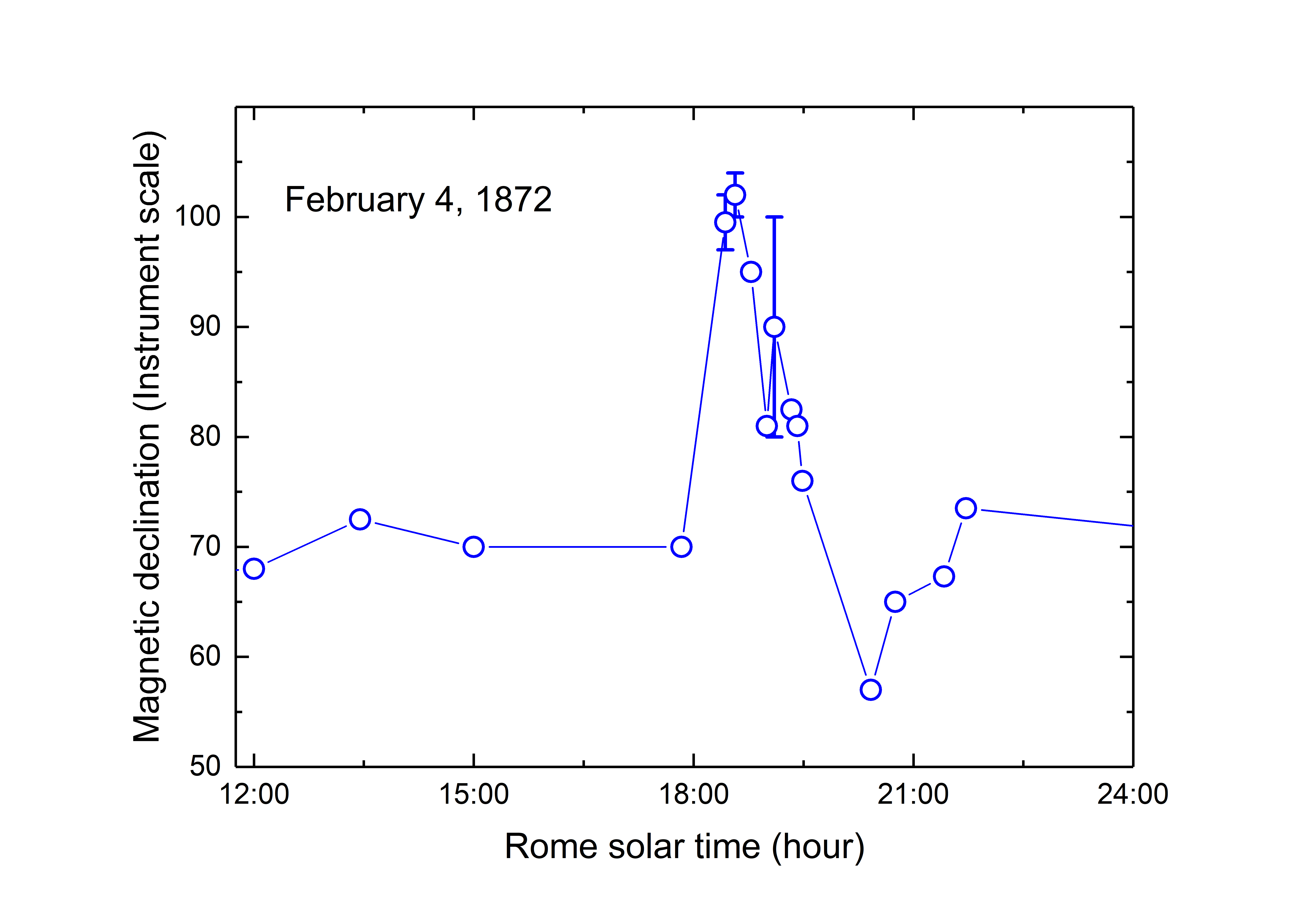}
\caption{The variations of the declination that were recorded at the laboratories of the Roman College on 4 February 1872 and reported in the Secchi's note \textit{Memoria sull’Aurora Elettrica del 4 Febbraio 1872}  \citep{Secchi1872}.
}
\label{declinometro}       			  
\end{figure}

In the note \citep{Secchi1872} it is reported that the magnetic axis of the aurora did not correspond to the magnetic meridian. 
The azimuth of the auroral centre, once taking into account the magnetic declination which was $13^o\,9'$, was measured to be over $10^o$ from the geographic meridian, that is, to $23^0$ from the magnetic meridian. The observations made in Livorno, Paris and Nimes were thus confirmed,
corroborating that the aurora was brighter in the east. Furthermore Secchi described the aurora as crown-like ovals, an extremely rare morphology in the low latitudes of Rome. The temporal evolution of this crown-like ovals is described extremely accurately, both in its privileged direction and in the rays that Secchi describes as \textit{parallel to the axis of the magnetic inclination needle, and the differences (which) show the extension of the perturbation actually suffered by the resultant of the magnetic forces}.\\
As already mentioned, the only instrument that was not out of scale during the entire event was the declinometer. However, it was possible to measure the decrease of the horizontal component of the magnetic field H during the peak of the event using the bifilar magnetometer, that was \textit{200 marks from the mean value, corresponding to a variation in the horizontal force of 0.0262}. This measure corresponds to a negative variation of H of about -600 nT, and thus near one fifth of the variation measured in Rome during the Carrington event in 1859 \citep{Boteler2006,Ptitsyna2012}.
This value is comparable with the -830 nT recorded in Colaba, Bombay during the same event \cite{Hayakawa2018}.
In Fig.~\ref{magnetic_table} and Fig.~\ref{declinometro} we show the magnetic observations reported in the Secchi's note. We note that the largest magnetic declination perturbations were observed around 18:30 (6:30~pm) Rome solar time (UT+0.83 hr), equivalent to 17:40 UT. This is in good accordance with the minimum of the H-component of the magnetic field as recorded in Colaba, Bombay at 22:30 Local Mumbai Time (LMT=UT+4.85 hr), i.e. 17:40 UT \citep{Hayakawa2018}.\\
Furthermore, Secchi reports that the maximum deviation measured by the declinometer corresponds to $1^o\,7'$, thus nearly one fourth of the deviation measured in Rome during the Carrington event \citep{Ptitsyna2012}.
Finally, the deviation of the vertical component of the magnetic field was measure at 19:40 Rome solar time, noticing an increase of $1^o\,11'$, a value that is similar in magnitude to the variation in direction of the horizontal component.
\subsection{Morphology and extent of the aurora}
As briefly discussed in the preceding paragraphs, the extent of the aurora was exceptional. Observations were in fact reported in France and England, but also in the South, in Africa and throughout Sicily, where it is reported that it was mistaken for an eruption of Etna. In the east it was observed as far as Istanbul, Suez, Cairo and Bombay. \\
The extreme extension of the phenomenon led Secchi to draw several
important conclusions. The first of these was that geographically distant observers did not see exactly the same aurora, but rather \textit{that everyone saw theirs}. Furthermore, he reported that the aurora formation altitude must be well above the earth's atmosphere, at extremely high altitudes.
Based on observations made simultaneously in Rome and Sondrio, by Prof. Lovisato, who described a morphology for the aurora with similar arc structures, he estimated the height of the aurora to be about 246 km. 
This estimate is in accordance with the mostly red appearance of the aurora and the consequent classification as type-d aurora \citep{Jones1971}, which has a typical height of 250 km.
He also wrote: \textit{since there is a contemporaneity of appearance in extremely distant sites, we must hypothesize for this phenomenon that the electric discharge that spreads from the atmosphere to the ground invests every observer, and that those rays that seem to us to converge in reality they are almost parallel to each other and to the resultant of the earth's magnetic forces. This idea is certainly not new, but now it seems no longer to be questioned by anyone}.\\ 
It must be reported that Secchi is also looking for a connection between aurora and meteorological conditions on a small geographical and temporal scale. This is probably due to his belief that there is a connection between meteorological, electrical and magnetic phenomena and solar events. In this regard Secchi wrote that the connection with the meteorological circumstances that preceded and followed the aurora were clear. Secchi wrote \textit{it is known that the aurora is a precursor of weather change in the polar regions; from what I saw in Rome I can assure you that this is the case}.\\
\subsection{Solar activity recorded in Rome and Moncalieri}
\begin{figure}
\centering
\includegraphics[width=14cm,height=12cm]{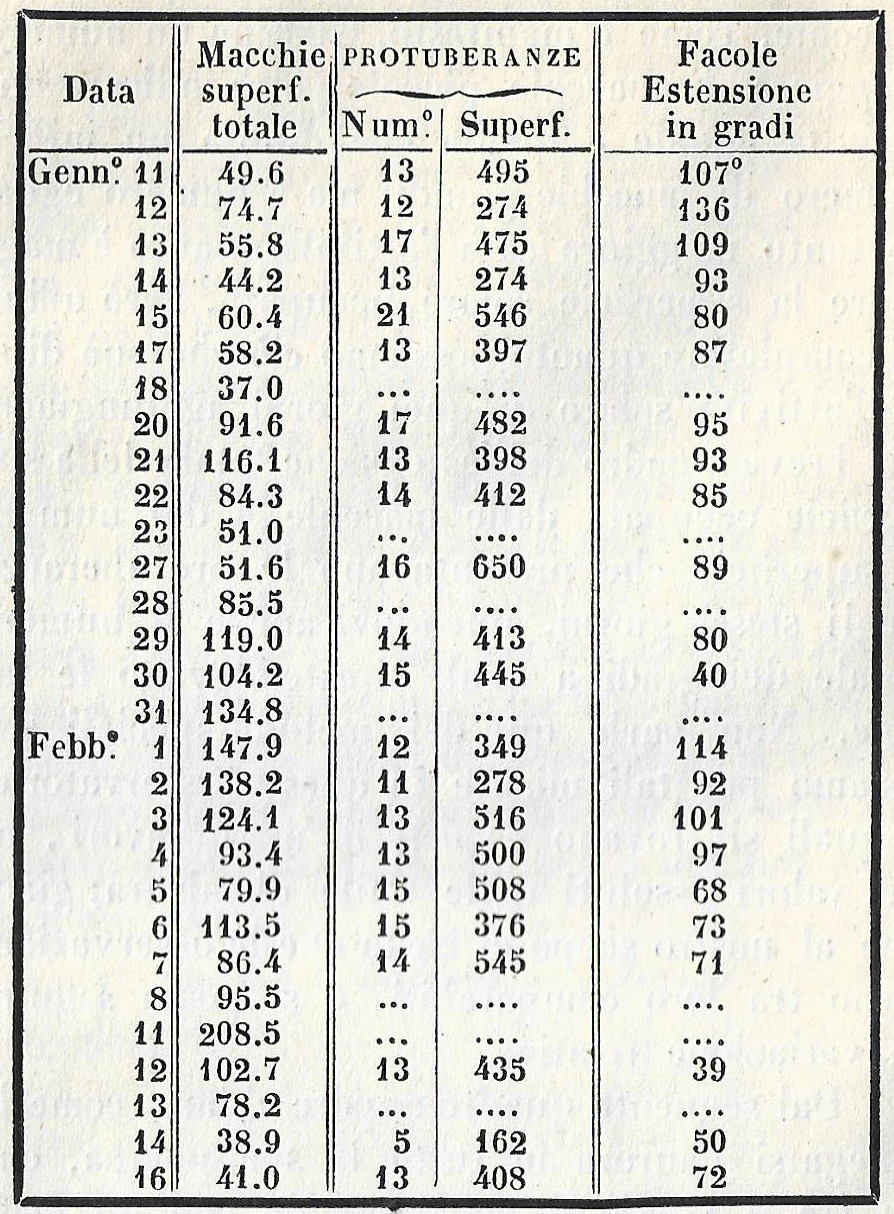}
\caption{Table of solar activity observations of January and February 1872 carried out at the Roman College. The different columns record: the date, the surface of solar sunspot (arbitrary units), the number and the extension of prominences, the extension of solar faculae. From the note \textit{Studii sull’Aurora Elettrica del 4 Febbraio 1872. G. Egidi} \citep{Egidi1872}.
}
\label{solar_Activity}       			  
\end{figure}

Solar activity was monitored on a daily basis at the Observatory of Roman College, as it was recently outlined by \cite{Carrasco2021}, where the authors provide a digitalized version of Secchi's prominence and sunspot observations for the period 1871–1875. Moreover, Angelo Secchi was in contact with colleagues across Italy and Europe to compare simultaneous observations. Therefore we have records of the observations made in January and February 1872 were we can search for a possible source of the great aurora of February 4.\\
Demonstrating once more its modern approach Secchi and the staff at the Observatory of Roman College records the surface extension of the sunspots observed in the solar photosphere. In Fig.~\ref{solar_Activity} we show the daily sunspots area as reported in a note on the aurora by one of the staff member at observatory \citep{Egidi1872}.
It is evident an enhanced activity in the days February 1 and 2, with an increment of $\sim30$\% that could be linked to the emergence of active regions that could be linked to flare and CME events.
This is also confirmed by the observations made by Francesco Denza in Moncaliera, one of Secchi's students and later director of the Vatican Observatory. He recorded an increment in the number of groups (g) and sunspots (s) as reported in \cite{Egidi1872}: from the baseline activity level of January 31 (g=6; s=87) the number of sunspots raised to a local maximum on February 2 (g=10; s=116) and then went back to lower levels on February 5 (g=8; s=98).

\subsection{Angelo Secchi's comments on the aurora}
Secchi concludes the note with a series of general considerations on the possible Sun-Earth connection. In particular, he points out that since the Carrington event of 1859 there is a suspicion that there is a relationship between auroral manifestations and solar activity, in particular with solar prominences and eruptions.\\
However, although a coincidence was noted between the two phenomena, it was evident that while many solar prominences were observed, no similar number of auroras were observed. In this regard, Secchi wrote: \emph{I do not intend to deny the known relationships between solar phenomena and auroras: but it is one thing to speak of an average coincidence during particular periods, it is quite another to expect that every single eruption produces an aurora or a magnetic perturbation [...] in fact, the first case may depend on a general state of the Sun which, by influencing the terrestrial globe through secondary causes, influences its magnetic state. The second case cannot be explained except by a direct transmission of
matter from the Sun to us or by means of an electric induction capable of being transmitted through the planetary vacuum. [...] in this regard there would be the possibility of a communication of a direct matter. However, this hypothesis, which is that of Mairan, is no longer happy, because the solar jets, observing them at the limb, are always at right angles with the vector ray from the Earth to the Sun, and the matter launched into space would not fall to the Earth, but it would go far from us. At most, the eruptions coming from the centre of the solar disk could reach the Earth, which however remain ordinarily invisible to us. In this case there would be a greater probability of coincidence with the sunspots and their formation, since these really seem to derive from the eruptions}.\\
The proposed analysis, despite being affected by the knowledge of the time, for example on the origin of sunspots that were supposed to be produced by the prominences, formulates correct hypotheses. For example, when he speculates the more central regions in the solar disk, associated with sunspots, as most likely to create the direct connection between the Sun and the Earth. Or when, citing the hypothesis of Dortous de Mairan  \citep{Mairan1731}, he introduces the concept of "solar matter" at the origin of the auroras. Hypothesis that the advent of the space missions and the discovery of the solar wind and CMEs will prove correct.
\section {Effects of the aurora on local and global terrestrial infrastructures}
As we briefly argued in the introduction, we consider the approach of Secchi's note absolutely modern and important, from a historical and scientific point of view, because Secchi, in addition to using multi-instrument observations thanks to the availability of the Collegio Romano observatory, became interested in the effects of the aurora on the observatory's telecommunication systems and describes communication effects on a global scale.\\
At around 8:00~pm on February 4, Secchi asked the Collegio's telegraph station for information to find out if \textit {extraordinary phenomena} had been observed. He was then informed that disturbances to the line had been reported from 5:45~pm and that the greatest disturbance occurred at 6:15~pm, in practice corresponding to the measurements of the maximum magnetic disturbance (Fig.~\ref{declinometro}).\\
Various effects were reported on US and European telegraph networks. Above all, disturbances were reported on telegraph networks, due to single-wire technology, which connected England with the islands in the English Channel \citep{Preece1872}.\\ 
But certainly, what made this event the first global event on a terrestrial scale was the disruption of the service connecting the United States and Europe and based on the transatlantic cable or \emph{canapo transatlantico\/} as Secchi calls it. 
In fact, Europe in the decade 1850-1860, had built about 100,000 km of telegraph lines.
While starting from 1866, after the short life of the first cable in 1858, two transoceanic cables were activated that connected the United States with Europe \citep[e.g.][]{Schwartz2008}.
It was therefore possible to send telegrams, at the current cost of about 9,000\$ per word, with a transmission speed of about 8 words per minute. However, this global cabling, initially made with a wire technology and which was laid at high latitudes, was obviously very sensitive to magnetic field variations and to what we now call geomagnetically induced currents (GICs).
\section {Conclusions}
In this paper we have briefly described Angelo Secchi's note \emph{Memoria sull’Aurora Elettrica del 4 Febbraio 1872} for the Notes of the Pontifical Academy of new Lincei on February 18, 1872 \citep{Secchi1872} on the great Aurora of February 4, 1872.\\
In our opinion, Secchi's note is extremely important for at least two reasons: $i)$ it is one of the first examples, if not the first in absolute, of the study of an auroral event based on simultaneous observations and coordinated by different instruments: telescopes, magnetometers, electrical and meteorological instruments; $ii)$ Secchi introduces in the same note the analysis of the effects of a great aurora on the technological systems, available at the time, both on a local and global scale. \\
This work was possible thanks to the availability, at the same observatory, of different instruments that were chosen and installed by Secchi as director and that allowed simultaneous measurements of different physical nature.
The idea of studying the contemporaneous effects that the magnetic storm has on ground infrastructures and including this analysis in the note, reveals the absolutely modern, and in some ways revolutionary for the time, vision of Secchi and his group of collaborators.\\
The nature of this class of physical processes, which today we call space weather events, is inherently complex and multi-scale. So, Secchi recognized almost 150 years ago, the need to deal with them in a systematic and multi-disciplinary way. This organization is also today at the basis of the study of heliophysical processes and their understanding. Furthermore, this is a valid approach nowadays if we want to predict the effects of possible severe space weather events on human technological infrastructures in space and on Earth. This long path has therefore been traced by scientists with a broad scientific vision, such as Angelo Secchi.\\
\begin{acknowledgements}
We thank the anonymous referees for their helpful suggestions. This research is partially supported by the Italian MIUR-PRIN \emph{ Circumterrestrial Environment: Impact of Sun-Earth Interaction\/} grant 2017APKP7T.
\end{acknowledgements}


\bibliography{Aurora1872}

\begin{thebibliography}{42}
\providecommand{\natexlab}[1]{#1}
\providecommand{\url}[1]{\texttt{#1}}
\providecommand{\urlprefix}{URL }
\providecommand{\eprint}[2][]{\url{#2}}

\bibitem[{{Alberti} et~al.(2018){Alberti}, {Consolini}, {De Michelis},
  {Laurenza}, and {Marcucci}}]{Alberti2018}
{Alberti}, T., G.~{Consolini}, P.~{De Michelis}, M.~{Laurenza}, and M.~F.
  {Marcucci}, 2018.
\newblock {On fast and slow Earth's magnetospheric dynamics during geomagnetic
  storms: a stochastic Langevin approach}.
\newblock \emph{Journal of Space Weather and Space Climate}, \textbf{8}, A56.
\newblock 10.1051/swsc/2018039.

\bibitem[{{Berrilli}(2020)}]{Berrilli2020}
{Berrilli}, F., 2020.
\newblock {Angelo Secchi e la nascita della meteorologia spaziale moderna}.
\newblock \emph{Quaderni di Storia della Fisica, Italian Physical Society},
  \textbf{1}, 33.
\newblock 10.1393/qsf/i2020-10073-6.

\bibitem[{{Berrilli} et~al.(2014){Berrilli}, {Casolino}, {Del Moro}, {Di Fino},
  {Larosa} et~al.}]{Berrilli2014}
{Berrilli}, F., M.~{Casolino}, D.~{Del Moro}, L.~{Di Fino}, M.~{Larosa},
  et~al., 2014.
\newblock {The relativistic solar particle event of May 17th, 2012 observed on
  board the International Space Station}.
\newblock \emph{Journal of Space Weather and Space Climate}, \textbf{4}, A16.
\newblock 10.1051/swsc/2014014.

\bibitem[{{Bigazzi} et~al.(2020){Bigazzi}, {Cauli}, and
  {Berrilli}}]{Bigazzi2020}
{Bigazzi}, A., C.~{Cauli}, and F.~{Berrilli}, 2020.
\newblock {Lower-thermosphere response to solar activity: an
  empirical-mode-decomposition analysis of GOCE 2009-2012 data}.
\newblock \emph{Annales Geophysicae}, \textbf{38}(3), 789--800.
\newblock 10.5194/angeo-38-789-2020.

\bibitem[{{Boteler}(2006)}]{Boteler2006}
{Boteler}, D.~H., 2006.
\newblock {The super storms of August/September 1859 and their effects on the
  telegraph system}.
\newblock \emph{Advances in Space Research}, \textbf{38}(2), 159--172.
\newblock 10.1016/j.asr.2006.01.013.

\bibitem[{{Brenni}(1993)}]{Brenni1993}
{Brenni}, P., 1993.
\newblock Il Meteorografo di Padre Angelo Secchi.
\newblock \emph{Nuncius}, \textbf{8}(1), 197 -- 247.
\newblock Https://doi.org/10.1163/182539183X00082,
  \urlprefix\url{https://brill.com/view/journals/nun/8/1/article-p197_8.xml}.

\bibitem[{{Cade} and {Chan-Park}(2015)}]{Cade2015}
{Cade}, W.~B., and C.~{Chan-Park}, 2015.
\newblock {The Origin of ``Space Weather''}.
\newblock \emph{Space Weather}, \textbf{13}(2), 99--103.
\newblock 10.1002/2014SW001141.

\bibitem[{{Carrasco} et~al.(2021){Carrasco}, {Nogales}, {Vaquero},
  {Chatzistergos}, and {Ermolli}}]{Carrasco2021}
{Carrasco}, V. M.~S., J.~M. {Nogales}, J.~M. {Vaquero}, T.~{Chatzistergos}, and
  I.~{Ermolli}, 2021.
\newblock {A note on the sunspot and prominence records made by Angelo Secchi
  during the period 1871-1875}.
\newblock \emph{Journal of Space Weather and Space Climate}, \textbf{11}, 51.
\newblock 10.1051/swsc/2021033.

\bibitem[{{Chinnici}(2017)}]{Chinnici2017}
{Chinnici}, I.
\newblock {The Maker and the Scientist: The Merz-Secchi Connection}, 39--68,
  2017.
\newblock 10.1007/978-3-319-41485-0\_3.

\bibitem[{{Chinnici} and {Consolmagno}(2021)}]{Chinnici2021}
{Chinnici}, I., and G.~{Consolmagno}, 2021.
\newblock {Angelo Secchi and Nineteenth Century Science. The Multidisciplinary
  Contributions of a Pioneer and Innovator}.
\newblock 10.1007/978-3-030-58384-2.

\bibitem[{{Di Fino} et~al.(2014){Di Fino}, {Zaconte}, {Stangalini}, {Sparvoli},
  {Picozza} et~al.}]{Difino2014}
{Di Fino}, L., V.~{Zaconte}, M.~{Stangalini}, R.~{Sparvoli}, P.~{Picozza},
  et~al., 2014.
\newblock {Solar particle event detected by ALTEA on board the International
  Space Station. The March 7th, 2012 X5.4 flare}.
\newblock \emph{Journal of Space Weather and Space Climate}, \textbf{4}, A19.
\newblock 10.1051/swsc/2014015.

\bibitem[{{Dortous de Mairan}(1731)}]{Mairan1731}
{Dortous de Mairan}, J.-J., 1731.
\newblock {Trait\'e physique et historique de l'aurore bor\'eale}.
\newblock M\'emoires de l'Acad\'emie Royale des Sciences, A Paris del
  l'Imprimerie Royale.

\bibitem[{{Egidi}(1872)}]{Egidi1872}
{Egidi}, G., 1872.
\newblock {Studii sull'aurora elettrica del 4 febbraio 1872}.
\newblock
  \urlprefix\url{https://drive.google.com/file/d/1Sc1d2CrDvEcgJd_FIn2OPBazSpjLNU5w/view?usp=sharing}.

\bibitem[{{Fawcett}(1872)}]{Fawcett1872}
{Fawcett}, T., 1872.
\newblock {The Aurora of February 4}.
\newblock \emph{Nature}, \textbf{5}(120), 302.
\newblock 10.1038/005302c0.

\bibitem[{{Hayakawa} et~al.(2018){Hayakawa}, {Ebihara}, {Willis}, {Hattori},
  {Giunta} et~al.}]{Hayakawa2018}
{Hayakawa}, H., Y.~{Ebihara}, D.~M. {Willis}, K.~{Hattori}, A.~S. {Giunta},
  et~al., 2018.
\newblock {The Great Space Weather Event during 1872 February Recorded in East
  Asia}.
\newblock \emph{\apj}, \textbf{862}(1), 15.
\newblock 10.3847/1538-4357/aaca40, \eprint{1807.05186}.

\bibitem[{{Hayakawa} et~al.(2019){Hayakawa}, {Ebihara}, {Willis}, {Toriumi},
  {Iju} et~al.}]{Hayakawa2019}
{Hayakawa}, H., Y.~{Ebihara}, D.~M. {Willis}, S.~{Toriumi}, T.~{Iju}, et~al.,
  2019.
\newblock {Temporal and Spatial Evolutions of a Large Sunspot Group and Great
  Auroral Storms Around the Carrington Event in 1859}.
\newblock \emph{Space Weather}, \textbf{17}(11), 1553--1569.
\newblock 10.1029/2019SW002269, \eprint{1908.10326}.

\bibitem[{{J.~M.}(1872)}]{JMH1872}
{J.~M.}, H., 1872.
\newblock {The Aurora of February 4}.
\newblock \emph{Nature}, \textbf{5}(123), 365.
\newblock 10.1038/005365a0.

\bibitem[{{Jones}(1971)}]{Jones1971}
{Jones}, A.~V., 1971.
\newblock {Auroral Spectroscopy}.
\newblock \emph{\ssr}, \textbf{11}(6), 776--826.
\newblock 10.1007/BF00216890.

\bibitem[{{Lanchester} et~al.(2009){Lanchester}, {Ashrafi}, and
  {Ivchenko}}]{Lanchester2009}
{Lanchester}, B.~S., M.~{Ashrafi}, and N.~{Ivchenko}, 2009.
\newblock {Simultaneous imaging of aurora on small scale in OI (777.4 nm) and
  N$_{2}$1P to estimate energy and flux of precipitation}.
\newblock \emph{Annales Geophysicae}, \textbf{27}(7), 2881--2891.
\newblock 10.5194/angeo-27-2881-2009.

\bibitem[{{Lilensten} and {Belehaki}(2009)}]{Lilensten2009}
{Lilensten}, J., and A.~{Belehaki}, 2009.
\newblock {Developing the scientific basis for monitoring, modelling and
  predicting space weather}.
\newblock \emph{Acta Geophysica}, \textbf{57}(1), 1--14.
\newblock 10.2478/s11600-008-0081-3.

\bibitem[{{Oliveira} et~al.(2020){Oliveira}, {Hayakawa}, {Bhaskar}, {Zesta},
  and {Vichare}}]{Oliveira2020}
{Oliveira}, D.~M., H.~{Hayakawa}, A.~{Bhaskar}, E.~{Zesta}, and G.~{Vichare},
  2020.
\newblock {A possible case of sporadic aurora observed at Rio de Janeiro}.
\newblock \emph{Earth, Planets and Space}, \textbf{72}(1), 82.
\newblock 10.1186/s40623-020-01208-z, \eprint{2005.13453}.

\bibitem[{{Orchiston}(2020)}]{Orchiston2020}
{Orchiston}, W., 2020.
\newblock {Book Review: Decoding the Stars: A Biography of Angelo Secchi,
  Jesuit and Scientist}.
\newblock \emph{Journal of Astronomical History and Heritage}, \textbf{23}(1),
  227--228.

\bibitem[{{Plainaki} et~al.(2020){Plainaki}, {Antonucci}, {Bemporad},
  {Berrilli}, {Bertucci} et~al.}]{Plainaki2020}
{Plainaki}, C., M.~{Antonucci}, A.~{Bemporad}, F.~{Berrilli}, B.~{Bertucci},
  et~al., 2020.
\newblock {Current state and perspectives of Space Weather science in Italy}.
\newblock \emph{Journal of Space Weather and Space Climate}, \textbf{10}, 6.
\newblock 10.1051/swsc/2020003.

\bibitem[{{Plainaki} et~al.(2016){Plainaki}, {Lilensten}, {Radioti},
  {Andriopoulou}, {Milillo} et~al.}]{Plainaki2016}
{Plainaki}, C., J.~{Lilensten}, A.~{Radioti}, M.~{Andriopoulou}, A.~{Milillo},
  et~al., 2016.
\newblock {Planetary space weather: scientific aspects and future
  perspectives}.
\newblock \emph{Journal of Space Weather and Space Climate}, \textbf{6}, A31.
\newblock 10.1051/swsc/2016024.

\bibitem[{{Preece}(1872)}]{Preece1872}
{Preece}, W.~H., 1872.
\newblock {Earth-Currents and the Aurora Borealis of February 4, 1872}.
\newblock \emph{\nat}, \textbf{5}(123), 368.
\newblock 10.1038/005368a0.

\bibitem[{{Ptitsyna} and {Altamore}(2012)}]{Ptitsyna2012}
{Ptitsyna}, N., and A.~{Altamore}, 2012.
\newblock {Father Secchi and the first Italian magnetic observatory}.
\newblock \emph{History of Geo- and Space Sciences}, \textbf{3}(1), 33--45.
\newblock 10.5194/hgss-3-33-2012.

\bibitem[{{Rigge}(1918)}]{Rigge1918}
{Rigge}, W.~F., 1918.
\newblock {Father Angelo Secchi}.
\newblock \emph{Popular Astronomy}, \textbf{26}, 589.

\bibitem[{Schwartz and Hayes(2008)}]{Schwartz2008}
Schwartz, M., and J.~Hayes, 2008.
\newblock A history of transatlantic cables.
\newblock \emph{IEEE Communications Magazine}, \textbf{46}(9), 42--48.
\newblock 10.1109/MCOM.2008.4623705.

\bibitem[{{Schwenn}(2006)}]{schwenn2006}
{Schwenn}, R., 2006.
\newblock {Space Weather: The Solar Perspectives}.
\newblock \emph{Living Rev. Sol. Phys.}, \textbf{3}, 2.
\newblock Https://doi.org/10.12942/lrsp-2006-2.

\bibitem[{{Secchi}(1872)}]{Secchi1872}
{Secchi}, A., 1872.
\newblock {Sull'aurora elettrica del 4 febbraio 1872}.
\newblock
  \urlprefix\url{https://drive.google.com/file/d/1qAVSQcxld7Pzv4I4d24Gv7H7oU1wqHH4/view?usp=sharing}.

\bibitem[{{Secchi}(1873)}]{SecchiProt1873}
{Secchi}, A., 1873.
\newblock {Sulla distribuzione delle protuberanze solari e loro relazione colle
  macchie coll' aggiunta di un riassunto de' labori spettroscopici fatti in
  questi ultimi anni all' osservatorio del Collegio romano.}

\bibitem[{{Secchi}(1875)}]{SecchiSoleil1875}
{Secchi}, A., 1875.
\newblock {Le Soleil}.
\newblock 10.3931/e-rara-14748.

\bibitem[{{Secchi}(1877)}]{Secchi1877}
{Secchi}, A., 1877.
\newblock {L'astronomia in Roma nel pontificato DI Pio IX.}

\bibitem[{{Silverman}(2006)}]{Silverman2006}
{Silverman}, S.~M., 2006.
\newblock {Comparison of the aurora of September 1/2, 1859 with other great
  auroras}.
\newblock \emph{Advances in Space Research}, \textbf{38}(2), 136--144.
\newblock 10.1016/j.asr.2005.03.157.

\bibitem[{{Silverman}(2008)}]{Silverman2008}
{Silverman}, S.~M., 2008.
\newblock {Low-latitude auroras: The great aurora of 4 February 1872}.
\newblock \emph{Journal of Atmospheric and Solar-Terrestrial Physics},
  \textbf{70}(10), 1301--1308.
\newblock 10.1016/j.jastp.2008.03.012.

\bibitem[{{Silverman} and {Cliver}(2001)}]{Silverman2001}
{Silverman}, S.~M., and E.~W. {Cliver}, 2001.
\newblock {Low-latitude auroras: the magnetic storm of 14-15 May 1921}.
\newblock \emph{Journal of Atmospheric and Solar-Terrestrial Physics},
  \textbf{63}(5), 523--535.
\newblock 10.1016/S1364-6826(00)00174-7.

\bibitem[{{Slatter}(1872)}]{Slatter1872}
{Slatter}, J., 1872.
\newblock {Aurora of Feb. 4, 1872}.
\newblock \emph{Monthly Notices of the Royal Astronomical Society},
  \textbf{32}, 317.
\newblock 10.1093/mnras/32.8.317.

\bibitem[{{Spogli} et~al.(2019){Spogli}, {Piersanti}, {Cesaroni}, {Materassi},
  {Cicone}, {Alfonsi}, {Romano}, and {Ezquer}}]{Spogli2019}
{Spogli}, L., M.~{Piersanti}, C.~{Cesaroni}, M.~{Materassi}, A.~{Cicone},
  L.~{Alfonsi}, V.~{Romano}, and R.~G. {Ezquer}, 2019.
\newblock {Role of the external drivers in the occurrence of low-latitude
  ionospheric scintillation revealed by multi-scale analysis}.
\newblock 10.1051/swsc/2019032.

\bibitem[{{Stone}(1872)}]{Stone1872}
{Stone}, E.~J., 1872.
\newblock {The Aurora of February 4}.
\newblock \emph{Nature}, \textbf{5}(127), 443.
\newblock 10.1038/005443b0.

\bibitem[{{Toynbee}(1873)}]{Toynbee1873}
{Toynbee}, C., H., 1873.
\newblock {Extract from log (2933) of ship `Cottica,' Captain D. F. M'Kechnie,
  containing a notice of the occurrence of corposants during the Aurora of
  February 4, 1872}.
\newblock \emph{Quarterly Journal of the Royal Meteorological Society},
  \textbf{1}(3), 96--96.
\newblock 10.1002/qj.4970010308.

\bibitem[{{Valach} et~al.(2019){Valach}, {Hejda}, {Revallo}, and
  {Bochn{\'\i}{\v{c}}ek}}]{Valach2019}
{Valach}, F., P.~{Hejda}, M.~{Revallo}, and J.~{Bochn{\'\i}{\v{c}}ek}, 2019.
\newblock {Possible role of auroral oval-related currents in two intense
  magnetic storms recorded by old mid-latitude observatories Clementinum and
  Greenwich}.
\newblock \emph{Journal of Space Weather and Space Climate}, \textbf{9}, A11.
\newblock 10.1051/swsc/2019008.

\bibitem[{{Ward} et~al.(2021){Ward}, {Sepp{\"a}l{\"a}}, {Yi{\v{g}}it},
  {Nakamura}, {Stolle} et~al.}]{Ward2021}
{Ward}, W., A.~{Sepp{\"a}l{\"a}}, E.~{Yi{\v{g}}it}, T.~{Nakamura}, C.~{Stolle},
  et~al., 2021.
\newblock {Role Of the Sun and the Middle atmosphere/thermosphere/ionosphere In
  Climate (ROSMIC): a retrospective and prospective view}.
\newblock \emph{Progress in Earth and Planetary Science}, \textbf{8}(1), 47.
\newblock 10.1186/s40645-021-00433-8.

\end{thebibliography}
   

\end{document}